\documentstyle[12pt]{amsart}

\def\bC{{\Bbb C}}

\def\bZ{{\Bbb Z}}

\def\cL{{\cal L}}

\def\cD{{\cal D}}
\def\cS{{\cal S}}
\def\cZ{{\cal Z}}

\newtheorem{thm}{Theorem}[section]
\newtheorem{lemma}[thm]{Lemma}
\newtheorem{prop}[thm]{Proposition}
\newtheorem{cor}[thm]{Corollary}
\newtheorem{defi}[thm]{Definition}
\newtheorem{rk}[thm]{Remark}
\newtheorem{cl}[thm]{Claim}

\def\prf{\noindent{\textsc{Proof}}\rm\ }
\def\endprf{\ \hfill $\Box$}

\newcommand{\be}{\begin{equation}}
\newcommand{\ee}{\end{equation}}
\newcommand{\bea}{\begin{eqnarray}}
\newcommand{\eea}{\end{eqnarray}}
\newcommand{\beas}{\begin{eqnarray*}}
\newcommand{\eeas}{\end{eqnarray*}}

\begin{document}

\baselineskip=16pt
\parindent=0pt

\title[Insertion-Elimination Lie algebra]{The structure of the Ladder Insertion-Elimination Lie algebra}

\author[Mencattini and Kreimer]{Igor Mencattini \and Dirk Kreimer$^\dagger$}

{ {\renewcommand{\thefootnote}{\fnsymbol{footnote}} \footnotetext{\kern-15.3pt $\dagger$ D.K.\ supported by CNRS; both authors supported in parts by NSF grant
DMS-0401262, Ctr.\ Math.\ Phys.\ at Boston Univ.; BUCMP/04-06.} }}

{ {\renewcommand{\thefootnote}{\fnsymbol{footnote}} \footnotetext{\kern-15.3pt {\bf AMS Subject Classification:} { 17B65, 17B70, 16W30, 81T18, 81T15.}} }}

{ {\renewcommand{\thefootnote}{\fnsymbol{footnote}} \footnotetext{\kern-15.3pt {\bf Keywords:} { Insertion-Elimination Lie algebras, Feynman graphs,
Dyson--Schwinger equations, Heisenberg algebra.}}}


\address{Boston University, Department of Mathematics and Statistics,
Boston University, 111 Cummington Street,
Boston, MA 02215, USA}

\email{igorre@@math.bu.edu}


\address{CNRS at IHES, 35,
route de Chartres, 91440, Bures-sur-Yvette, France}

\email{kreimer@@ihes.fr}



\date{}
\maketitle

\begin{abstract} We continue our investigation into the insertion-elimination Lie algebra $\cL_{L}$ of Feynman graphs in the ladder case,
emphasizing the structure of this Lie algebra relevant for future applications in the study of Dyson--Schwinger equations. We work out
the relation to the classical infinite dimensional Lie algebra ${\frak gl}_{+}(\infty)$ and we determine the cohomology of $\cL_{L}$.
\end{abstract}

\vspace{1cm}

\section{Introduction}
In the last few years perturbative QFT has been shown to have a rich algebraic structure \cite{Kr} leading to relations with apparently unrelated sectors of
mathematics like non-commutative geometry and Riemann-Hilbert like problems \cite{C-K 1, C-K 2}. Such extraordinary relations can be resumed, to some extent, by
the existence of a commutative, non co-commutative Hopf algebra $\cal H$ defined on the set of Feynman diagrams.

We will continue the investigation started in \cite{MK} where we discussed first relations of perturbative QFT with the representation theory of Lie algebras. In
that paper we introduced the ladder Insertion-Elimination Lie algebra $\cL_{L}$ and we discussed relations of this Lie algebra with some more classical (infinite
dimensonal) Lie algebras.

In what follows we describe in greater detail the  structure of this Insertion-Elimination Lie algebra. The plan of the paper is as follows: we will start the
present paper in section two by some motivation for the relevance of the ladder insertion elimination Lie algebra $\cL_{L}$ for full  QFT. In particular we
stress the relation of $\cL_{L}$  with the  quantum equations of motion or Dyson-Schwinger equations (DSEs).

In section three we recollect some basic fact about the Lie algebra $\cL_{L}$ taken from \cite{MK}.

 Section four and section five are the core of this paper:
 in section four we give a structure theorem that stresses the relation of the Lie algebra $\cL_{L}$ with the classical infinite dimensional Lie algebra
${\frak gl}_{+}(\infty)$.

Finally, in section five we collect some basic result about the cohomology of the Lie algebra $\cL_{L}$.
\section{The significance of $Z_{n,m}$}
The Lie algebra $\cL_{L}$ on generators $Z_{n,m}$ is an insertion elimination Lie algebra \cite{MK} gained from these operations applied to a cocommutative and
commutative Hopf algebra $H_{\rm comm}$ built on generators (ladders) $t_n$, $n\geq 0$, $\Delta(t_n)=\sum_{j=0}^n t_j\otimes t_{n-j}$, on which it acts as a
derivation
$$Z_{i,j}(t_n)=\Theta(n-j)t_{n-j+i}, $$
where $\Theta(n-j)$ is defined as $\Theta(n-j)=1$ for $n-j\geq 0$, and $0$ otherwise.\\
This seems to give just a glimmer of the full insertion elimination Lie algebra of \cite{C-K}, which acts as a derivation on the full Hopf algebra of Feynman
graphs in a renormalizable quantum field theory.

Nevertheless, a full understanding of $Z_{n,m}$ goes a long way in understanding the full insertion elimination Lie algebra \cite{LL2004}, using the fact that
$\cL_{L}$ acts on elements in the full Hopf algebra which are homogenous in the appropriate grading resulting from the Hochschild cohomology of that very Hopf
algebra \cite{annals}.

There are two powerful reasons for that: i) quantum field theory sums over all skeleton graphs in a symmetric fashion, ii) non-linear Dyson--Schwinger equations
(DSEs) modify linear DSEs precisely by the anomalies generated by a non-vanishing $\beta$-function. The first fact ensures that we can work on homogenous
elements in the Hopf algebra, the second one ensures that there are effective methods available to deal with the operadic aspects of graph insertions.

Here, DSEs are introduced combinatorially via a fixpoint equation in the Hochschild cohomology of a connected graded commutative Hopf algebra. Let us summarize
the main features which emerged in recent work \cite{Houches,annals,polylog,LL2004}. Under the Feynman rules the Hochschild one-cocycles provided by the Hopf
algebra of graphs  map to integral operators provided by the underlying skeletons of the theory. Renormalization conditions are determined by suitable boundary
conditions for the integral equations so generated. The DSEs determine the Green functions from this Hochschild cohomology of the Hopf algebra of Feynman graphs,
which is itself derived from free quantum field theory and the choice of renormalizable interactions.

Indeed, following \cite{polylog,annals} the identification of these one cocycles  leads to a combinatorial Dyson--Schwinger equation: \bea \Gamma^{\underline{r}}
& = & 1 + \sum_{{p\in H_L^{[1]}} \atop {{\rm res}(p)
  =  \underline{r}}} \frac{\alpha^{\vert p\vert}}{{\rm Sym}(p)}
B_+^p(X_{p})\nonumber\\ & = & 1+\sum_{{\Gamma\in H_L}\atop{ {\rm res}(\Gamma)=\underline{r}}}\frac{\alpha^{\vert\Gamma\vert}\Gamma}{{\rm
Sym}(\Gamma)}\;,\label{DSE}\eea
 where the first sum is over a finite (or
countable) set of Hopf algebra primitives $p$, Feynman graphs such that \be \Delta(p)=p\otimes e + e \otimes p, \end{equation}  indexing the closed Hochschild
one-cocycles $B_+^{p}$ above, while the second sum is over all one-particle irreducible graphs contributing to the desired Green function, all weighted by their
symmetry factors. Here, $X_{p}$ is a polynomial in all $\Gamma^{\underline{r}}$, and the superscript $\underline{r}$ ranges over the finite set (in a
renormalizable theory) of superficially divergent Green functions. It indicates the number and type of external legs reflecting the monomials in the underlying
Lagrangian. We use ${\rm res}(p)=\underline{r}$ to indicate that the external legs of the graph $p$ are of type $\underline{r}$. The structure of these equations
allows for a proof of locality using Hochschild cohomology \cite{annals}, which is also evident using a coordinate space approach \cite{Bergbau}.

These fixpoint equations are solved by an Ansatz
\begin{equation} \Gamma^{\underline{r}}=1+\sum_{k=1}^\infty \alpha^k c^{\underline{r}}_k .\end{equation}
We grant ourselves the freedom to call such an equation a DSE or a combinatorial equation of motion for a simple reason: the DSEs of any renormalizable quantum
field theory can be cast into this form. Crucially, in the above it can be shown (see \cite{LL2004}, which we follow here)  that \be X_{p}=\Gamma^{{\rm res}(p)}
X_{\rm coupl}^{\mid p \mid},\end{equation} where $X_{\rm coupl}$ is a connected Green functions which maps to an invariant charge under the Feynman rules. This
is rather obvious: consider, as an example, the vertex function in quantum electrodynamics: a $n$ loop primitive graph $p$ contributing to it provides $2n+1$
internal vertices, $2n$ internal fermion propagators and $n$ internal photon propagators. An invariant charge \cite{Gross} is provided by a vertex function
multiplied by the squareroot of the photon propagator and the fermion propagator. Thus the integral kernel corresponding to $p$ is dressed by $2n$ invariant
charges, and one vertex function. This is a general fact: each integral kernel corresponding to a Green function with external legs $\underline{r}$ in a
renormalizable quantum field theory is dressed by a suitable power of invariant charges proportional to the grading of that kernel, and one additional apperance
of $\Gamma^{\underline{r}}$ itself. This immediately shows that for a vanishing $\beta$-function the DSEs are reduced to a linear set of equations, and that the
general case can be most efficiently handled by an expansion in the breaking of conformal symmetry induced by a non-vanishing $\beta$-function. Thus, a complete
understanding of the linear case goes a long way in understanding the full solution. This emphasizes the crucial role which the insertion-elimination Lie algebra
\cite{C-K} in the ladder case \cite{MK} plays in the full theory: it defines an algebra of graphs which provide an underlying field of residues which is then
extended by the contributions resulting from a non-trivial $\beta$-function: the resulting scaling anomalies extend the Hopf algebra of graphs to a
non-cocommutative one, at the same time they force the appearance of new transcendental numbers and result in the appearance of non-trivial representations of
the symmetric group in the operad of graph insertions \cite{LL2004}. Here, we study the underlying linear DSEs which would suffice for a vanishing
$\beta$-function.

Indeed, we now define the linear DSE associated to the system above: \bea \Gamma_{\rm lin}^{\underline{r}} & = &  1 + \sum_{{p\in H_L^{[1]}} \atop {{\rm res}(p)
  =  \underline{r}}} \frac{\alpha^{\vert p\vert}}{{\rm Sym}(p)}
B_+^p(\Gamma_{\rm lin}^{\underline{r}}).\label{DSEl}\eea The Hochschild closedness of $B_+^p$ then ensures that we obtain a Hopf algebra isomorphic to the word
Hopf algebra based on letters $p$, which we obtain as the underlying Dyson skeletons in the expansion of $\Gamma^{\underline{r}}$.

The solutions of the linear DSE above are graded by the order in $\alpha$ and by the augmentation degree \be \Gamma_{\rm lin}^{\underline{r}}=1+\sum_{j=1}^\infty
\alpha^j c_j =1+\sum_{j=1}^\infty d_j.\end{equation} Here, $c_j$ is the sum of all words of order $\alpha^j$, where the degree $\mid\! w\!\mid$ of a word $w$ is
the sum of the degree of its letters, and the degree of a letter is the loop number of the accompanying skeleton graph. These words uniquely correspond to
Feynman graphs obtained by inserting primitive graphs into each other, where insertion now happens at a single vertex or edge in accordance with that linear DSE.

On the other hand, $d_j$ is the sum \be d_j=\sum_{w\in H_{\rm aug}^j } \alpha^{\mid w\mid}w,\end{equation} of all words made out of $j$ letters, and we set
$\mid\! w\!\mid_{\rm aug}=j$, the augmentation degree.

Having defined the associated linear system, the propagator-coupling dualities \cite{BK} provide the general solution once the representation theory of the
symmetric group has been established, which reflects the operadic nature of graph insertions \cite{LL2004}. But a complete understanding of linear
Dyson--Schwinger equations comes first. To this end, it is profitable to study the action of the insertion elimination Lie algebra which acts on the Hopf algebra
of graphs in that case.

In this paper, we start some groundwork by clarifying the structure of the insertion elimination Lie algebra which relates to the Hopf algebra structure of a
linear DSE. The crucial point is always the identification of the Hochschild closed one-cocycles in the Hopf algebra of graphs $B_+^p$, typically parametrized by
primitive elements $p$ of the Hopf algebra of graphs \cite{annals}.

We first mention that the Hopf algebra of graphs contains, as a corollary of the results in \cite{C-K}, a sub Hopf algebra $H_w$ of graphs generated by the
linear DSE. It is naturally based on graphs which can be regarded as words, with corresponding insertion-elimination Lie algebra $L_w$. It acts on the Hopf
algebra $H_w$ as \be Z_{w_1,w_2}(w)=\begin{cases} w_1v \;\;\text{if}\;\; w=w_2v \;\;\text{for some}\;\; v\\ 0,\;\; \text{if $w$ has not this form.}
\end{cases}\end{equation} The Lie bracket in $L_w$ is then
 \bea [Z_{w_1,w_2},Z_{w_3,w_4}] & = & Z_{\overline{Z_{w_1,w_2}(w_3)},w_4}
-Z_{w_3,\overline{Z_{w_2,w_1}(w_4)}}\nonumber\\ & & -Z_{\overline{Z_{w_3,w_4}(w_1)},w_2}+Z_{w_1,\overline{Z_{w_4,w_3}(w_2)}}\nonumber\\
 & & -\delta^K_{w_2,w_3}Z_{w_1,w_4} +\delta^K_{w_1,w_4}Z_{w_3,w_2}.
 \eea
See \cite{C-K} for notation.

The significance of the Lie algebra $Z_{n,m}$ comes from the fact that the map $B_+=\sum_p\alpha^{\mid p\mid} B_+^p$ maps the linear DSE to the fundamental DSE
(which also underlies the polylog \cite{polylog}) \be X=1+B_+(X),\end{equation} where $B_+$ is of order $\alpha$. Note that $B_+$ is not homogenous in $\alpha$
(there are primitive graphs of any degree in the coupling), but it is homogenous in the augmentation degree: all terms in its defining sum enhance this degree by
one.

There is a natural inclusion $\iota_H$ from $H_{\rm comm}$ to $H_w$ which sends $t_n\to d_n$. This induces a map \be \iota_L:\; L\to L_w,\; Z_{n,m}\to \sum_{\mid
w_1\mid_{\rm aug}=n,\mid w_2\mid_{\rm aug}=m}\frac{Z_{w_1,w_2}}{\#(m)},\end{equation} where $\#(m)$ is the number of words of degree $m$, such that
$\iota_H(Z_{n,m}(t_k))=\iota_L(Z_{n,m})(\iota(t_k))$. It is compatible with the Lie bracket: \be
[\iota_L(Z_{n_1,m_1}),\iota_L(Z_{n_2,m_2})]\left(\iota_H(t_n)\right)=\iota_H\left([Z_{n_1,m_1},Z_{n_2,m_2}](t_n)\right).\end{equation} As long as we study linear
DSEs, the ladder insertion elimination Lie algebra on generators $Z_{n,m}$ suffices, where it now acts by increasing and decreasing the augmentation degree. In
\cite{LL2004} the reader can find a discussion of the Galois theory which is missing to handle the general case. The study of such questions in QFT is a
beautiful mathematical problem in its own right. It gives mathematical justification to early ideas \cite{MacTod} of the use of anomalous dimensions and
bootstrap equations in QFT to absorb short-distance singularities. Progress along these lines following \cite{polylog, LL2004} will be reported in future work.
 We now continue to treat $\cL_{L}$.

\section{Generalities about the Ladder Insertion-Elimination Lie algebra.}
Let us recall some basic definition from \cite{MK} to which we refer for the details omitted in what follows.

Let us introduce the Lie algebra $\cL_{L}$ via generators and relations:
\begin{defi}
$\cL_{L}=\hspace{4pt}{\text {span}}_{\bC}<Z_{n,m}\vert\hspace{3pt}n,m\in\bZ_{\geq 0}>,$
with:
\begin{eqnarray} \big[Z_{n,m}, Z_{l,s}\big] & = &
\Theta(l-m)Z_{l-m+n, s}-\Theta(s-n)Z_{l,s-n+m}\nonumber\\ & & -
\Theta(n-s)Z_{n-s+l,m}+\Theta(m-l)Z_{n,m-l+s}\nonumber\\ & &
-\delta_{m,l}Z_{n,s}+\delta_{n,s}Z_{l,m},\label{e1}
\end{eqnarray}
where:
\begin{equation}
\begin{cases}
\Theta(l-m) &=0 \hspace{5pt} \text{if $l<m$}, \\
\Theta(l-m) &=1 \hspace{5pt} \text{if $l\geq m$}
\end{cases}
\end{equation}
and where $\delta_{n,m}$ is the usual Kronecker delta:
\begin{equation}
\begin{cases}
\delta_{n,m} &=1 \hspace{5pt} \text{if $m=n$},\\
\delta_{n,m} &=0 \hspace{5pt} \text{if $n\neq m$}.
\end{cases}
\end{equation}
\end{defi}

We start with the following:
\begin{cor}\cite{MK}
1) $\cL_{L}$ is $\bZ$-graded Lie algebra:
$$\cL_{L}=\oplus_{i\in\bZ}l_i$$
where each for each $Z_{n,m}\in l_i$, $deg(Z_{n,m})=i=n-m$ and ${\text {dim}}_{\bC}\hspace{4pt}l_{i}=+\infty$;\\ 2) $\cL$ has the following decomposition:
$${\cal L}_{L}= L^{+}\oplus L^{0}\oplus L^{-};$$
where $L^{+}=\oplus_{n>0}l_n$, $L^{-}=\oplus_{n<0}l_n$ and $L^{0}=l_0$.
\end{cor}
\prf
The statements follow from the definition of graded Lie algebra, i.e $\cL_{L}$ is $G$-graded (where $G$ is any abelian group) if $\cL_{L}=\oplus_{i\in G}l_i$ and $[l_i,l_j]\subset l_{i+j}$, and from the formula (\ref{e1}).

\endprf


We conclude this section with the following:

\begin{prop}\label{p1} Each element $Z_{n,m}\in\cL_{L}$ can be written in the following form:

\begin{equation}
Z_{n,m}=[Z_{n,0},Z_{0,m}]+\Theta(n-m)Z_{n-m,0}+\Theta(m-n)Z_{0,m-n}-\delta_{n-m,0}Z_{0,0}.\label{e}
\end{equation}

\end{prop}

\prf The statement follows trivially applying formula (\ref{e1}) to the elements $Z_{n,0}$ and $Z_{0,m}$ in the cases $n>m$, $n<m$ and $n=m$.
\endprf

\begin{rk}
The previous proposition is equivalent to the following (vector space) decomposition of the Lie algebra $\cL_{L}$:
$${\cL}_{L}=[{\cD},{\cD}]\oplus{\cD};$$
where we defined:
$${\cD}={\frak a}_{+}\oplus {\frak a}_{-}\oplus{\bC}$$
and ${\frak a}_{+}=span_{\bC}<Z_{n,0}:n>0>$, ${\frak a}_{-}=span_{\bC}<Z_{0,n}:n>0>$ and $\bC$ is the trivial Lie algebra generated by $Z_{0,0}$. In fact ${\frak a}_{+}$ and ${\frak a}_{-}$ are commutative sub algebras of $\cL_{L}$ and $Z_{0,0}$ is a central element.
\end{rk}

\section{Structure of the Lie algebra $\cL_{L}$.}
Let us start this section with two statements whose proofs are collected at the end of the this section.

\begin{thm}\label{T1} The center of the Lie algebra $\cL_{L}$ has dimension one and it is generated by the
element $Z_{0,0}$.
\end{thm}

\begin{thm}\label{T2}
$l^{0}$ is a maximal abelian sub-algebra of $\cL_{L}$.
\end{thm}

In what follows we will show that the Lie algebra $\cL_{L}$ is not simple. Let us introduce the following:

\begin{defi}\label{D}\cite{MK}
$${\frak gl}_{+}(\infty)={\text {span}}_{\bC}<E_{i,j}:Z_{i,j}-Z_{i+1,j+1}\vert\hspace{3pt}i,j\in\bZ_{\geq 0}>.$$
\end{defi}

We have:

\begin{prop}
1) $[E_{i,j}, E_{r,k}]=E_{i,k}\delta_{j,r}-E_{r,j}\delta_{k,i}$;\\
2) $\frak gl_{+}(\infty)$ is an ideal in $\cL_{L}$.
\end{prop}
\prf The proof of 1) and 2) is a simple but tedious application of the commutator formula (\ref{e1}).

\endprf

We can then define the quotient Lie algebra:
$$ C=\cL_{L}/{\frak gl}_{+}(\infty)$$
and consider the exact sequence:
\begin{equation}
0 @>>> {\frak gl}_{+}(\infty) @>>> {\cL}_{L} @>\pi>> {C} @>>> 0.\label{es}
\end{equation}
We want to understand the Lie algebra $C$. To accomplish this goal we need the following
\begin{prop}\label{p}
$${\frak gl_{+}(\infty)}=[{\cL}_{L}, {\cL}_{L}].$$
\end{prop}

\prf
Let us prove the two inclusions.\\
${\frak gl}_{+}(\infty)\subset [{\cL}_{L}, {\cL}_{L}]$ since
from the definition of $\frak gl_{+}(\infty)$:
$$E_{i,j}=Z_{i,j}-Z_{i+1,j+1}=[Z_{i,0},Z_{0,j}]-[Z_{i+1,0},Z_{0,j+1}],$$
where the second equality follows from the formula (\ref{e}).\\
To show the other inclusion, i.e $[{\cL}_{L}, {\cL}_{L}]\subset{\frak gl}_{+}(\infty)$, it suffices to observe that for any two generators $Z_{h,p}$ and
$Z_{r,q}$ of $\cL_{L}$ we have that their commutator is given by the difference between two elements having same degree (see formula (\ref{e1})), say $Z_{n,m}$ and $Z_{l,s}$
with $n-m=l-s$.

Under the hypothesis that $k=n-m=l-s>0$ and that $s>m$ (the other cases are completely analogous) we can write their difference as follows: \beas Z_{n,m}-Z_{l,s}
&  = & Z_{m+k,m}-Z_{s+k,s}=Z_{m+k,m}-Z_{m+k+1,m+1}+\\
& & +Z_{m+k+1,m+1}-....-Z_{s+k-1,s-1}+Z_{s+k-1,s-1}-Z_{s+k,s},\eeas
which expresses the difference between $Z_{n,m}$ and $Z_{l,s}$ as finite linear combination
of elements in ${\frak gl}_{+}(\infty)$.
\endprf

In particular we can rephrase the previous proposition in the following way:
\begin{lemma}\label{c}

Two generators $Z_{n,m}$ and $Z_{l,s}$ are $\frak gl_{+}(\infty)$-equivalent if and only if they have the same
degree, i.e:
$$Z_{n,m}\sim Z_{l,s}\iff\hspace{4pt} deg(Z_{m,n})=deg(Z_{l,s}).$$
\end{lemma}

\prf
If $Z_{n,m}$ and $Z_{l,s}$ are have the same degree then they are equivalent by the argument used to prove the proposition above.\\
Suppose now that the difference between $Z_{n,m}$ and $Z_{l,s}$ can be written as a (finite) linear combination of elements in $\frak gl_{+}(\infty)$ and also
that $n-m\neq l-s$ (w.l.o.g.\ we can assume that $n-m>0$ and that $l-s>0$).

Under these assumptions and from formula (\ref{e}), it follows also that: $Z_{n-m,0}-Z_{l-s,0}=\sum_{finite}a_{i}E_{p_i,q_i}$ But this has as consequence that each of
these two elements are finite linear combinations of (homogeneous) elements in $\frak gl_{+}(\infty)$, so that we can write:
$Z_{n-m,0}=\sum_{finite}c_{i}E_{r_i,k_i}$ and $Z_{l-s,0}=\sum_{finite}c_{i}E_{t_i,v_i}$. Rewriting the right hand side of each of those two equalities in terms
of the generators $Z_{n,m}$, it is clear that such equations can not hold.
\endprf

From the proposition \ref{p} it follows that $C$ is a (maximal) commutative Lie algebra coming as quotient of $\cL_{L}$.\\
Let us now introduce a set of (natural) generators for $C$. Since the set $$<Z_{n,m}\vert\hspace{4pt} n,m\in\bZ_{\geq 0}>$$ is a basis for $\cL_{L}$ and since
$\pi:{\cL}_{L}\longrightarrow C$ is a surjection, we have that:
$$<{\overline Z}_{n,m}=\pi(Z_{n,m})\vert\hspace{4pt} n,m\in\bZ_{\geq 0}>$$
is a set of generators for $C$. Moreover it follows from the lemma \ref{c} that for $n>m$ we have $Z_{n,m}\sim Z_{n-m,0}$, for $m>n$ we have $Z_{n,m}\sim
Z_{0,m-n}$ and for $n=m$ we have $Z_{n,m}\sim Z_{0,0}$. So defining $Z_n={\overline Z}_{n,0}$, $Z_{-n}={\overline Z}_{0,n}$ for $n>0$ and $Z_0={\overline
Z}_{0,0}$, we get
$$C=span_{\bC}<Z_{n}\vert n\in\bZ>.$$
The fact that such elements are also linearly independent (i.e they form a basis for $C$) follows also from the lemma \ref{c}.

In what follows in this section we want to prove that:

\begin{thm}\label{?}
The exact sequence (\ref{es}) does not split, i.e the Lie algebra $\cL_{L}$ is not the semi-direct product of the Lie algebra $\frak gl_{+}(\infty)$ with the (commutative) Lie algebra $C$.
\end{thm}

Before addressing the proof of theorem \ref{?} we need to introduce some preliminaries to make the paper as self-contained as possible.

The exact sequence (\ref{es}) implies that the Lie algebra $\cL_{L}$ is a non-abelian extension of the commutative Lie algebra $C$ by the Lie algebra $\frak
gl_{+}(\infty)$ (for generalities about non-abelian extension of Lie algebras we refer to the paper \cite{AMR} and references therein). If $\frak g$, $\frak h$ and $\frak e$ are Lie algebras:

\begin{defi}\cite{AMR}
We will say that the Lie algebra $\frak e$ is an extension of the Lie algebra $\frak g$ by the Lie algebra $\frak h$ if $\frak g$, $\frak h$ and $\frak e$ fit
in the following exact sequence:
\begin{equation}
0@>>> {\frak h} @>>> {\frak e} @>\pi>> {\frak g} @>>> 0,\label{es2}
\end{equation}
Moreover we will say that two such extensions $\frak e$ and $\frak e'$ are equivalent if and only if $\frak e$ and $\frak e'$ are isomorphic as Lie algebras.
\end{defi}
Let $Der(\frak h)$ be the Lie algebra of derivations of $\frak h$, $\alpha',\alpha\in{\text Hom}_{\bC}({\frak g},{\text Der}({\frak h}))$ and
$\rho',\rho\in{\text Hom}_{\bC}(\Lambda^2{\frak g},{\frak h})$. On the set of the couples $(\alpha,\rho)$ introduced above we define the equivalence relation:
$$(\alpha,\rho)\sim(\alpha',\rho')\iff\exists\hspace{4pt} b\in Hom_{\bC}({\frak g},{\frak h})$$
such that:
$$\alpha'(x).\xi=\alpha(x).\xi+[b(x),\xi]_{\frak h},$$
$$\rho'(x,y)=\rho(x,y)+\alpha(x).b(y)-\alpha(y).b(x)-b([x,y]_{\frak g})+[b(x),b(y)]_{\frak h}.$$

Then we have the following:

\begin{thm}\label{??}\cite{AMR}
1) The classes of isomorphism of the extensions of the Lie algebra $\frak g$ by the Lie algebra $\frak h$ given by the exact sequence (\ref{es2}), are in one-to-one correspondence
with the classes of equivalence $[(\alpha, \rho)]$ such that:
$$[\alpha(x),\alpha(y)]_{Der(\frak h)}.{\xi}-\alpha([x,y]_{\frak g}).{\xi}=[\rho(x\wedge y),\xi]_{\frak h};$$
$$\sum_{{\rm cyclic}}\big(\alpha(x).\rho(y,z)-\rho([x,y]_{\frak g},z)\big)=0;$$
for every $x,y,z\in{\frak g}$ and $\xi\in\frak h$.\\
2) The Lie algebra structure induced on the vector space ${\frak e}={\frak h}\oplus{\frak g}$ by the datum $(\alpha,\rho)$ is given by:
\begin{equation}
[({\xi}_1,x_{1}),({\xi}_2,x_{2})]{\frak e}=([\xi_1,\xi_2]_{\frak h}+\alpha(x_1).\xi_2-\alpha(x_2).\xi_1+\rho(x_1,x_2),[x_1,x_2]_{\frak g})\label{p2}.
\end{equation}
\end{thm}

We apply this result to our setting, where ${\frak g}=C$ and ${\frak h}={\frak gl}_{+}(\infty)$.

The exact sequence (\ref{es}) tells us that we have:
$${\cL}_{L}\simeq {\frak gl}_{+}(\infty)\oplus{C}$$
where such splitting holds in the category of vector spaces. We first show the following:

\begin{prop}\label{p3}
The Lie algebra structure on $\cL_{L}$ given by the bracket (\ref{e1}) corresponds to the couple $(\alpha,\rho)$ defined by:

$$\alpha(Z_{n}).(E_{i,j})=\Theta(n)\sum_{k\geq 0}(E_{n+k,j}\delta_{i,k}-E_{i,k}\delta_{n+k,j})+$$
$$\Theta(-n)\sum_{k\geq 0}(E_{k,j}\delta_{k+n,i}-E_{i,k+n}\delta_{j,k})$$
for $n\neq 0$ and
$$\alpha(Z_0)\equiv 0;$$
while:

$$\rho(Z_{n},Z_{m})=0$$

if $n,m\geq 0$ or $n,m\leq 0$ and

$$\rho(Z_{n},Z_{-m})=\sum_{k=0}^{m-1}E_{n-m+k,k}$$

if $n>m$, and:

$$\rho(Z_{n},Z_{-m})=\sum_{k=0}^{m-1}E_{k,m-n+k},$$

if $n<m$.

\end{prop}
\prf The proof follows comparing formula $(\ref{e1})$ with formula (\ref{p2}).
\endprf

We now remark that:

\begin{lemma}\label{!}\cite{AMR}
Given:
\begin{equation}
0@>>> {\frak h} @>>> {\frak e} @>\pi>> {\frak g} @>>> 0,\label{es3}
\end{equation}
as in (\ref{es2}), any splitting $s:{\frak g}\longrightarrow {\frak e}$ (at the vector space level) of the previous exact sequence, induces a map $\alpha_s\in Hom_{\bC}\big({\frak g}, Der({\frak h})\big)$, via the following:
$$\alpha_{s}(X).\xi=[s(X),\xi];$$
for each $X\in \frak g$ and each $\xi\in\frak h$.
\end{lemma}

\begin{prop}
The map $\alpha\in{\text Hom}_{\bC}\big(C,{\text Der}({\frak gl}_{+}(\infty))\big)$, defined in proposition, \ref{p3} is induced by the linear map $s:C\longrightarrow {\cL}_{L}$, where:

\begin{equation}
s(Z_{n})=\Theta(n)Z_{n,0}+\Theta(-n)Z_{0,n}-\delta_{n,0}Z_{0,0}.\label{s}
\end{equation}

\end{prop}
\prf
The map $s$ defined in formula (\ref{s}) is a section of the projection $\pi:{\cL_{L}}
\longrightarrow C$ defined by the exact sequence (\ref{es}), i.e $s\in Hom_{\bC}(C,{\cL_{L}})$ and $s\circ\pi=Id_{C}$. From the lemma (\ref{!}) induces a linear map:
$$\alpha_{s}:C\longrightarrow Der({\frak gl_{+}(\infty)}),$$
defined by:
$$\alpha_{s}(x).\xi=[s(x),\xi]_{\cL_{L}}.$$
It is now easy to check that this map is the same defined in the proposition \ref{p3}.
\endprf

\vspace{20pt}

We are now almost ready to prove theorem \ref{?}. We only need to remark the following: from theorem \ref{??} we have that a given extension $(\alpha,\rho)$
of the Lie algebra $C$ by the Lie algebra $\frak gl_{+}(\infty)$ will split, i.e will be equivalent to a semi-direct product of the these two Lie algebras, if and
only if $(\alpha,\rho)\sim(\alpha',0)$ i.e if and only if $\alpha'$ is a morphism of Lie algebras. From the theorem \ref{??} this is equivalent to ask for the existence
of a linear map $b:C\longrightarrow {\frak gl_{+}(\infty)}$ such that $s+b:C\longrightarrow {\cL_{L}}$ is a morphism of Lie algebras. Note that we are working
in the category of graded Lie algebras so the requirement for such map is that it has to preserve the grading. Showing that such a map does not exist we will be able to conclude that the exact sequence does not split.

\prf(Theorem \ref{?}) Suppose we can define a linear map $b:C\longrightarrow \frak gl_{+}(\infty)$ such that $s+b:C\longrightarrow \cL_{L}$ is a morphism of (graded)
Lie algebras. That means that we can find elements $\sum_{i=1}^{M}a_{h_{i}}E_{h_i+1,h_i}\in \frak gl_{+}(\infty)$ and
$\sum_{i=1}^{N}b_{k_{j}}E_{k_j,k_j+1}\in\frak gl_{+}(\infty)$ such that $b(Z_1)=\sum_{i=1}^{M}a_{h_{i}}E_{h_i+1,h_i}$,
$b(Z_{-1})=\sum_{i=1}^{N}b_{k_{j}}E_{k_j,k_j+1}$ and furthermore
$$0=[(s+b)(Z_1),(s+b)Z_{-1}]=[Z_{1,0}+\sum_{i=1}^{M}a_{h_{i}}E_{h_i+1,h_i},Z_{0,1}+\sum_{j=1}^{N}b_{k_{j}}E_{k_j,k_j+1}].$$

We can calculate such a commutator by re-writing each of the terms $E_{i,j}$ in the sums in terms of the generators $Z_{n,m}$, and applying to such terms the
brackets given in formula (\ref{e1}). The result, written in terms of the generators $E_{i,j}$, takes the form:
$$-E_{0,0}+\sum_{j=1}^{N}b_{k_{j}}(E_{k_{j}+1,k_{j}+1}-E_{k_j,k_j})+\sum_{i=1}^{M}a_{h_i}(1+b_{h_{i}})(E_{h_{i}+1,h_{i}+1}-E_{h_i,h_i})=0.$$

The right hand side us of the previuos sum can be reorganized in term of the summands $E_{j+1,j+1}-E_{j,j}$ as follows:
$$-\sum_{i\geq 0}^{L}\phi_j (E_{j+1,j+1}-E_{j,j}),$$
for $L$ is the biggest between $N$ and $M$ and the $\phi_j$'s are coefficients.\\
So that we have that:
$$E_{0,0}=\sum_{i\geq 0}^{L}\phi_j (E_{j+1,j+1}-E_{j,j})=-\phi_{0}E_{0,0}+\sum_{j\geq 0}(\phi_{j+1}-\phi_{j})E_{j,j}+\phi_{L}E_{L+1,L+1},$$
that clearly give us a contradiction.
\endprf

\subsection*{Proof of Theorems \ref{T1} and \ref{T2}}
We now conclude this section by giving the proofs for theorems \ref{T1} and \ref{T2}.\\

We recall from \cite{MK} that the Lie algebra $\cL_{L}$ has an obvious module:
\begin{defi}\cite{MK}
$${\cal S}=\bigoplus_{n\geq 0}{\bC}t_n={\bC}[t_0,t_1,t_2,t_3.....].$$
We will assign a degree equal to $k$ to the generator $t_{k}$ for each $k\geq 0$. ${\cal L}_{L}$ acts on $\cal S$ via the following:
\begin{gather}
Z_{n,m}t_k=0\hspace{5pt}if\hspace{5pt} m>k,\notag\\
Z_{n,m}t_k=t_{k-m+n}\hspace{5pt}if\hspace{5pt} m\leq k\label{e0}.
\end{gather}
\end{defi}
In what follows we will indicate by $\cZ(\cL_{L})$ the center of the Lie algebra $\cL_{L}$.\\
\noindent \prf(Theorem \ref{T1}). It is obvious that $\bC Z_{0,0}\subset\cZ(\cL_{L})$. Let us prove the other inclusion. Let us suppose that there is some element
$\alpha\in\cL_{L}$, not proportional to $Z_{0,0}$ and that belongs to the center of $\cL_{L}$. W.l.o.g.\ we assume
\begin{equation}
\alpha=\sum_{i=1}^{k}a_i Z_{n_i,m_i}=\sum_{i:\, n_i, m_i\neq 0}b_i Z_{n_i,m_i}+\sum_{i:\, {\tilde n}_i\neq 0}c_i Z_{{\tilde n}_i,0}+\sum_{i:\ {\tilde m}_i\neq
0}d_i Z_{0,{\tilde m}_i},\label{al}
\end{equation}
where all the ${\tilde n}_i$'s (${\tilde m}_i$'s) are different from 0 and ${\tilde n}_i\neq {\tilde n}_j$ (${\tilde m}_i\neq {\tilde m}_j$)
if $i\neq j$, and $(n_i,m_i)\neq (n_j, m_j)$ if $i\neq j$.\\
We will prove that $\alpha$ defined above is equal to zero by showing that the coefficients $b_i, c_i$ and $d_i$ are all equal to zero.
We will split the proof of this assertion into two lemmas.\\
\begin{lemma}\label{l1}
If $\alpha\in\cZ(\cL_{L})$, $\alpha$ defined as above, then $b_i=d_i=0$  for each $i$.
\end{lemma}
\prf
Let us consider some element $Z_{n,0}\in \cL_{L}$ such that $0<n\leq\hspace{3pt}min_{i}\{m_i,{\tilde m}_i\}$. Then using formula (\ref{e1}), we get:
$$[Z_{n,0},\alpha]=\sum_i b_i[Z_{n,0},Z_{n_i,m_i}]+\sum_i d_i[Z_{n,0},Z_{0,{\tilde m}_i}]=$$
$$\sum_i b_i(Z_{n_i+n,m_i}-Z_{n_i,m_i-n})+\sum_{i}d_i(Z_{n,{\tilde m}_i}-Z_{0,{\tilde m}_i-n}).$$
Note the all the ${\tilde m}_i$'s are different (and different from $0$), while in the set of the $m_i$'s (also all different from $0$) we can have repetitions.

Let us now define the set $M\doteq \{m_1,....,m_k,{\tilde m_1},...,{\tilde m_r}\}$ and let us consider the disjoint union:
$$M=M_{1}\cup \ldots \cup M_{s},$$
where each $M_i$ corresponds to the set of all indices in $M$ which are equal to some given index $l_i$, say. We remark  once more that for each $i$
$M_i\cap\{{\tilde m_1},\cdots ,{\tilde m_r}\}$ contains at most one element since in the set $\{{\tilde m_1},...,{\tilde m_r}\}$ we do not have repetitions.\\
Now let us consider $p_1=l_1-n$ ($\geq 0$, by the condition we imposed on $n$), and the corresponding element $t_{p_1}\in\cS$.
Since $\alpha$ belongs to $\cZ(\cL_{L})$ and since $n>0$, we have:

\begin{equation}
0=[Z_{n,0},\alpha](t_{p_1})=-\Big(\sum_{{i:\, m_i\in M_1}}b_it_{p_1-m_i+n+n_i}+\sum_{{i:\, {\tilde m}_i\in M_1}}d_it_{p_1-{\tilde m}_i+n}\Big).\label{f}
\end{equation}
\begin{rk}
We observe that all the indices in $M_1$ are equal to $l_1$ and that $p_1=l_1-n$. Moreover the $n_i$'s in the first sum of the right hand side in formula (\ref{f})
are all different (since by assumption we have that $(n_i,m_i)\neq (n_j,m_j)$
unless $i=j$ and in our case all the $m_i$ belong to the class $M_1$).
Finally we notice that the last sum, if not equal to zero, contains only one term.\\
\end{rk}
Let us now suppose that $M_1\cap \{m_1,\cdots ,m_k\}$ and $M_1\cap\{{\tilde m_1},\cdots ,{\tilde m_r}\}$ are both not empty (the cases where one of those intersections or both of them are empty are completely analogous). From the previuos remark it follows that
$$0=[Z_{n,0},\alpha](t_{p_1})=-\big(\sum_{i:\,m_i\in M_1}b_it_{l_1-n-l_1+n+n_i}+\sum_{i:\,{\tilde m}_i\in M_1}d_it_{l_1-n-l_1+n}\big)=$$
$$-\big(\sum_i b_it_{n_i}+d_1t_{0}\big).$$
Now since all the $n_i$ in the first sum are different we have that $d_1=0$ and all $b_i=0$.\\
Now using the same argument we can proceed with the sets $M_2$,...,$M_s$, to show that each of the coefficients $b_i$ and $c_i$ are equal to $0$.
\endprf

From the lemma \ref{l1} we conclude that if $\alpha\in\cZ(\cL_{L})$, $\alpha$ defined as in equation (\ref{al}), then:
$$\alpha=\sum_{i}c_iZ_{n_i,0}.$$
To conclude the proof of the theorem (\ref{T1}) we have to show that:
\begin{lemma}
If $\alpha\in\cZ(\cL_{L})$ and $\alpha=\sum_{i}c_iZ_{n_i,0},$ then $c_i=0$ for each $i$.
\end{lemma}
\prf
We first notice that we can suppose all $n_i\neq 0$ and $n_1<n_2...$. Let us now consider some element $Z_{0,n}$, such that $n\geq{\text {max}}_{i}\{n_i\}$. Since we suppose $\alpha=\sum_{i}c_iZ_{n_i,0}$ to be in the center of $\cL_{L}$, we can write:
$$0=[\alpha, Z_{0,n}]=\sum_{i}c_i[Z_{n_i,0},Z_{0,n}]=\sum_i c_i(Z_{n_i,n}-Z_{0,n-n_i}).$$
By the hypothesis on $n$ and on the $n_i$'s we conclude that all the $c_i$'s are equal to zero.
\endprf

\vspace{20pt}
\prf (Theorem \ref{T2}). Let us suppose that $l^0$ is not maximal abelian sub-algebra of $\cL_{L}$,
i.e that there exists $\cL_{L}\ni\alpha\notin l^0$, $\alpha=\sum_{i=1}^{n}a_iZ_{n_i,m_i}$, such that:
$$[\alpha, Z_{k,k}]=0,\hspace{4pt}\forall\hspace{4pt} k>0.$$
Without loss of generality we can suppose that in each of $(n_i,m_i)$'s, $n_i\neq m_i$ (if
no, $\alpha=\beta +\sum_{i}f_iZ_{n_i,n_i}$ and $[\beta,
Z_{k,k}]=[\alpha, Z_{k,k}]$).

Such an element can be written as:

\begin{equation}
\alpha=\sum_{{i:\, m_{i}\neq 0,\, n_{i}\neq 0}}b_iZ_{n_i,m_i}+\sum_{{i:\, {\tilde n}_i\neq 0}}c_iZ_{{\tilde n}_i,0}+
\sum_{{i:\, {\tilde m}_{i}\neq 0}}d_iZ_{0,{\tilde m}_i}.\label{al2}
\end{equation}

\begin{rk}
We note that in formula (\ref{al2}) all the $n_i$'s and the $m_i$'s are different from
$0$ and also that ${\tilde n}_{i}\neq {\tilde n}_{j}$ and ${\tilde m}_{i}\neq
{\tilde m}_{j}$ for each $i\neq j$.
\end{rk}
We will prove that such element is identically equal to zero, showing that each of the coefficients in the equation (\ref{al2}) is equal to zero. We will divide the proof of this statement in two lemmas.\\
\begin{lemma}
Given $\alpha\in l^0$, defined as in formula (\ref{al2}), we have that $c_i=d_i=0$ for all $i$.
\end{lemma}
\prf
Let us fix integer $k$, $0<k\leq {\text {min}}_{i}\hspace{3pt}\{n_i,m_i,{\tilde n}_i, {\tilde m_i}\}$. Then we get:
$$[\alpha, Z_{k,k}]=\sum_{i:\,m_{i}\neq 0, n_{i}\neq 0}b_i[Z_{n_i,m_i},Z_{k,k}]+\sum_{i:\,{\tilde n}_i\neq 0}c_i[Z_{{\tilde n}_i,0},Z_{k,k}]+\sum_{i:\,{\tilde m}_{i}\neq 0}d_i[Z_{0,{\tilde m}_i},Z_{k,k}]=$$
$$\sum_{i:\,{\tilde n}_i\neq 0}c_i(Z_{k+{\tilde n}_i,k}-Z_{{\tilde n}_i,0})+\sum_{i:\,{\tilde m}_{i}\neq 0}d_i(Z_{0,{\tilde m}_i}-Z_{k,k+{\tilde m}_i}),$$
since:
$$[Z_{n_i,m_i},Z_{k,k}]=0,\hspace{4pt}\forall\hspace{4pt}\{n_i,m_i\}\hspace{4pt}{\text {such that}}\hspace{4pt} n_i\geq k,\hspace{3pt} m_i\geq k,$$

$$[Z_{{\tilde n}_i,0},Z_{k,k}]=Z_{k+{\tilde n}_i,k}-Z_{{\tilde n}_i,0}\hspace{4pt}{\text {if}}\hspace{4pt}0<k\leq {\tilde n}_i,\hspace{4pt}{\text {and}}$$

$$[Z_{0,{\tilde m}_i},Z_{k,k}]=-Z_{k,{\tilde m}_i+k}+Z_{0,{\tilde m}_i}\hspace{4pt}{\text {if}}\hspace{4pt}0<k\leq {\tilde m}_i.$$

Since $\alpha$ commutes with all the elements of the sub-algebra $l^0$, we have:
$$0=\sum_{i:\,{\tilde n}_i\neq 0}c_i(Z_{k+{\tilde n}_i,k}-Z_{{\tilde n}_i,0})+\sum_{i:\,{\tilde m}_{i}\neq 0}d_i(Z_{0,{\tilde m}_i}-Z_{k,k+{\tilde m}_i}).$$
But in the right hand side of the previous formula the first sum contains only elements of positive degree while the second sum contains only those of negative
degree, thus the sum is equal to zero if and only if separately
$$\sum_{i:\,{\tilde n}_i\neq 0}c_i(Z_{k+{\tilde n}_i,k}-Z_{{\tilde n}_i,0})=0\hspace{5pt}{\text
{and}}\hspace{4pt}\sum_{i:\,{\tilde m}_{i}\neq 0}d_i(Z_{0,{\tilde m}_i}-Z_{k,k+{\tilde m}_i})=0.$$ From this it follows that all $c_i$'s and $d_i$'s are equal to
zero. Indeed, consider the sum containing the $c_i$'s (the one contains the $d_i$'s can be treated in the same way):
$$\sum_{i:\,{\tilde n}_i\neq 0}c_i(Z_{k+{\tilde n}_i,k}-Z_{{\tilde n}_i,0})=0.$$
Since $k\neq 0$ and since ${\tilde n}_i\neq {\tilde n}_j$ if $i\neq j$, all the elements $Z_{k+{\tilde n}_i,k}-Z_{{\tilde n}_i,0}$ are linearly independent.
\endprf

Summarizing, so far we have proved that if a given element $\alpha$ commutes with each of the elements in $l^0$, then:
\begin{equation}
\alpha=\sum_{i:\,n_i\neq 0, m_i\neq 0}b_iZ_{n_i, m_i}.\label{al3}
\end{equation}
\begin{lemma}
If $\big[\alpha,l^0\big]=0$, with $\alpha$ defined as in (\ref{al3}), then all the $b_i$'s are equal to $0$.
\end{lemma}
\prf
Let us decompose the element $\alpha$ in term of elements of positive and negative degree, i.e:
$$\alpha=\sum_{i}a_iZ_{n_i,m_i}=\sum_{j}\big(\sum_{i\geq 0}b_iZ_{r_i+s_j,r_i}\big)+\sum_{j}\big(\sum_{i\geq 0}c_iZ_{p_i,p_i+t_j}\big),$$
\begin{rk}
We remark that in $\alpha$ elements of the same (negative or positive) degree could be present; as an example of such element (of positive degree) we can consider:
$$\beta_j=\sum_ib_iZ_{r_i+s_j,r_i},\hspace{4pt}{\text {for a given}}\hspace{4pt}j\hspace{4pt}$$
or the element (of negative degree):
$$\gamma_j=\sum_ic_iZ_{p_i,p_i+t_j},\hspace{4pt}{\text {for a given}}\hspace{4pt}j\hspace{4pt}.$$
\end{rk}
From the previuos remark let us re-write $\alpha$ as:
$$\alpha=\sum_j\beta_j+\sum_j\gamma_j,$$
each $\beta_j\in L^+$ and each $\gamma_j\in L^-$.

Let us now consider some element $Z_{k,k}\in l^0$ and let us take the commutator of such element with $\alpha$
$$[\alpha,Z_{k,k}]=\sum_{j}[\beta_j,Z_{k,k}]+\sum_j[\gamma_j,Z_{k,k}].$$
Since $\cL_{L}$ is a graded Lie algebra and since ${\text {deg}}\hspace{4pt}Z_{k,k}=0$, we have that
$${\text {deg}}\hspace{4pt}[\beta_j,Z_{k,k}]=s_j\hspace{4pt},\hspace{3pt}\forall{j}$$
and similarly
$${\text {deg}}\hspace{4pt}[\gamma_j,Z_{k,k}]=-t_j\hspace{4pt},\hspace{3pt}\forall{j}.$$
Hence
$$[\alpha, Z_{k,k}]=0\iff[\beta_j,Z_{k,k}]=0\hspace{4pt}{\text {and}}\hspace{4pt}[\gamma_j,Z_{k,k}]=0\hspace{4pt},\hspace{3pt}\forall j.$$
We are left to prove that any homogeneous element commuting with all the elements in $l^0$ can not exist.

So, to fix ideas, let us now consider some element of positive degree $s$, say, $\beta=\sum_{i=1}^{l}a_iZ_{n_i+s,n_i}$ and let us suppose that
\begin{equation}
[\beta, Z_{k,k}]=0\hspace{4pt}\forall\hspace{4pt}k\geq 1.\label{beta}
\end{equation}
Without loss of generality we can further assume that $0<n_1<n_2<...<n_k$ (that $\beta$ fulfills the hypothesis is constrained by the assumptions given for the
element $\alpha$ defined in formula (\ref{al3}), which translates for $\beta$ into the condition $n_i\neq 0$). To conclude, it suffices to show that each of the $a_i$'s
of $\beta=\sum_{i=1}^{l}a_iZ_{n_i+s,n_i}$ is equal to $0$. So let us consider $k=n_2$ in formula (\ref{beta}). Applying the formula (\ref{e1}) to this case, we get:
$$[\beta,Z_{k,k}]=a_1[Z_{n_1+s,n_1},Z_{n_2,n_2}]+\sum_{i\geq 2}a_i[Z_{n_i+s,n_i},Z_{n_2,n_2}]=$$
$$=a_1\big(Z_{n_2+s,n_2}-\Theta(n_2-n_1-s)Z_{n_2,n_2-s}-\Theta(n_1+s-n_2)Z_{n_1+s,n_1}+\delta_{n_1+s,n_2}Z_{n_2,n_1}\big),$$
since $\sum_{i\geq 2}a_i[Z_{n_i+s,n_i},Z_{n_2,n_2}]=0$.\\
By the previous formula and the hypothesis for the $n_i$'s, we conclude that $[\beta,Z_{n_2,n_2}]=0\iff\hspace{4pt}a_1=0$. Taking $k=n_3, n_4,....$, and using
the same argument, we can conclude that each of the $a_i$'s is equal to zero.
\endprf
\section{Cohomology of the Lie algebra $\cL_{L}$.}

In what follows we will describe in some details the cohomology of  the  Lie algebra $\cL_{L}$. We will start with an explicit calculation for the dimension of
the first cohomology group (with trivial coefficients) and we will continue using the general machinery to calculate the higher cohomology groups.

Let us introduce on the Lie algebra $\cL_{L}$ the derivation $Y$ via the following:
\begin{equation}
Y.Z_{n,m}\equiv [Y,Z_{n,m}]=(n-m)Z_{n,m}.\label{e2}
\end{equation}
Let now consider the extension of the Lie algebra $\cL_{L}$ obtained by adding to $\cL_{L}$ the derivation $Y$.

\begin{defi}

$${\check\cL_{L}}=span_{\bC}<Z_{n,m}, Y\vert n,m\in{\bZ}_{\geq 0}>,$$

where the commutator $[Z_{n,m},Z_{l,s}]$ is given by formula (\ref{e1}) and $[Y,Z_{n,m}]=(n-m)Z_{n,m}.$

\end{defi}

\vspace{10pt}

We have the following:

\begin{thm}\label{T1x}
$$1.\hspace{4pt}{\text {dim}}_{\bC}\hspace{4pt}H^{1}(\check\cL_{L},\bC)=1;$$

$$2.\hspace{4pt}{\text {dim}}_{\bC}\hspace{4pt}H^{1}(\cL_{L},\bC)=+\infty.$$

\end{thm}
\prf  The elements of $H^{1}(\cL_{L},\bC)$ are in one-to-one correspondence with the elements $\phi\in\hspace{4pt}{\text {Hom}}_{\bC}(\cL_{L},\bC)$
such that:
$$\phi([Z_{n,m},Z_{l,s}])=0,$$
for each $Z_{n,m},Z_{l,s}\in\cL_{L}$. As a consequence of proposition \ref{p1}, equation (\ref{e}),
we have that the value of $\phi$ on a given element $Z_{n,m}$, depends only on the degree of such an element. In fact, given:

$$Z_{n,m}=[Z_{n,0},Z_{0,m}]+\Theta(n-m)Z_{n-m,0}+\Theta(m-n)Z_{0,m-n}-\delta_{n-m,0}Z_{0,0},$$

$$\phi(Z_{n,m})=\Theta(n-m)\phi(Z_{n-m,0})+\Theta(m-n)\phi(Z_{0,m-n})-\delta_{n-m,0}\phi(Z_{0,0}).$$
From this remark, it follows that ${\text {dim}}_{\bC}\hspace{4pt}H^{1}(\cL_{L},\bC)=+\infty$.\\
This proves the second assertion. To show that also the first holds, let us observe that, since:
$$0=\phi([Y,Z_{n,m}])=(n-m)\phi(Z_{n,m}),$$
for each $Z_{n,m}\in \cL_{L}$, then $\phi(Z_{n,m})=0$ for any element $Z_{n,m}$ with degree different from zero, so that we can write:
$$\phi(Z_{n,n})=c_{\phi}(n)\in\bC.$$
On the other hand, since the value of $\phi$ on a given element depends only on the degree of such an element, we have that:
$$\phi(Z_{n,m})=c_{\phi}\delta_{n-m,0},$$
or, in other words:
$$H^1(\check\cL_{L},\bC)\simeq\bC.$$

\endprf

\vspace{20pt}

To go to the higher cohomology groups we need to introduce some notation and some (classical) results about the cohomology of Lie algebras. Let us start with the
following result.

\begin{lemma}\label{l}
Let ${\frak gl}(n)$ the (Lie) algebra of $n\times n$ matrices (with entries in $\bC$). Let us define the direct system of (Lie) algebras:

\begin{equation}
 ...\rightarrow{\frak gl}(n-1)\rightarrow{\frak gl}(n)\rightarrow{\frak gl}(n+1)\rightarrow...\label{S}
\end{equation}
where the arrows are given by the standard inclusions, i.e $A\in{\frak gl}(n)$ is mapped into ${\tilde A}\in{\frak gl}(n+1)$ such that ${\tilde
a}_{i,j}=a_{i,j}$, $\forall i\leq n,\hspace{3pt}j\leq n$ and ${\tilde a}_{i,j}=0$ if $i>n$ or $j>n$. Then the direct limit of such a direct system is isomorphic
to the Lie algebra $\frak gl_{+}(\infty)$ introduced in definition \ref{D}.

\end{lemma}
\prf

The proof follows immediately from the definition \ref{D}.

\endprf

Let us next quote a result about the cohomology of the Lie algebra of the general linear group ${\frak gl}(n)$ \cite{F}:

\begin{thm}\label{th}
{\it 1)}. The cohomology ring of the Lie algebra ${\frak gl}(n)$ is an exterior algebra in $n$ generators of degree $1,3,...,2n-1$:
$$H^{\bullet}({\frak gl}(n))=\Lambda [c_{1},c_{3},....., c_{2n-1}];$$
{\it 2)} for any given $n$, the (inclusion) map, defined in formula (\ref{S}), lemma \ref{l}:
$$i:{\frak gl}(n)\longrightarrow {\frak gl}(n-1)$$
induce a map $i^{\ast}$ in cohomology:
$$i^{\ast}:H^{\bullet}({\frak gl}(n))\longrightarrow H^{\bullet}({\frak gl}(n-1)),$$
such that:
$$i^{\ast}:H^{p}({\frak gl}(n))\longrightarrow H^{p}({\frak gl}(n-1))$$
is an isomorphism for $p<n$, and it maps to zero the top degree generator when $p=n$;

\noindent {\it 3)} from 1), 2) and the previous lemma \ref{l}, it follows that the cohomology ring of the Lie algebra $\frak gl_{+}(\infty)$ is a (non finitely generated)
exterior algebra having generators only in odd degree:
$$H^{\bullet}({\frak gl}_{+}(\infty))=\Lambda [c_{1},c_{3},.....].$$
\end{thm}

\prf

We refer to the book \cite{F} for the proof of 1) and 2). Part 3) follows from lemma (\ref{l}) where we have identified the direct limit of the Lie algebras ${\frak gl}(n)$ with the Lie algebra
$\frak gl_{+}(\infty)$.
\endprf

Now let us go back to the exact sequence (\ref{es}). This induces the following exact sequence in cohomology (\cite{F}, \cite{W}): \bea & &  0@>>> H^{1}(C)
@>f^{1}>>H^{1}({\cL}_{L}) @>>> H^{1}({\frak gl}_{+}(\infty))^{C} @>{\delta}>> H^{2}(C)\nonumber\\ & & @>f^{2}>>H^{2}({\cL}_{L}) @>>> H^{2}({\frak
gl}_{+}(\infty))^{C} @>{\delta}>>\cdot\cdot\cdot .\label{es1} \eea Here, by $H^{i}({\frak gl}_{+}(\infty))^C$ we understand the (sub)-vector space of
$C$-invariant elements in $H^{i}({\frak gl}_{+}(\infty))$, i.e the space
  $\{a\in H^{i}({\frak gl}_{+}(\infty))\vert x.a=0,\hspace{5pt} \forall\hspace{3pt} x\in C\}$.
Using theorem (\ref{th}) and the previous exact sequence we have one of our main results.

\begin{thm}

The cohomology groups $H^{i}({\cL}_{L})$ are infinite dimensional. In particular the Lie algebra $\cL_{L}$ has infinite many non equivalent central extensions.

\end{thm}

\prf
The theorem follows easily analysing the following segment of the exact sequence (\ref{es1}):

\begin{equation}
\cdot\cdot\cdot @>>> H^{p}({\frak gl}_{+}(\infty))^{C} @>{\delta}>> H^{p+1}(C)@>f^{p+1}>>H^{p+1}({\cL}_{L}) @>>> H^{p+1}({\frak gl}_{+}(\infty))^{C}@>>>\cdot\cdot\cdot
\end{equation}

From this we see that for $p$ odd $f^{p+1}$ is surjective as $H^{p+1}({\frak gl}_{+}(\infty))=0$ (by Thm.\ref{th}). On the other hand dim $H^{p}({\frak
gl}_{+}(\infty))<\infty$, so that ${\text{ker}}(\hspace{3pt}f)$ is finite dimensional.

 We can argue in an analogous way in the case of $p$ even; in this case we
use the fact that the map $f^{p+1}$ is injective as $H^{p}({\frak gl}_{+}(\infty))=0$.

 The statement about central extensions follows now from the fact that
those are in one-to-one correspondence with the elements of the group $H^{2}({\cL}_{L})$.
\endprf

We want to make the statement in the previous theorem about the central extensions of the Lie algebra $\cL_{L}$ more precise. In particular we have that:
\begin{prop}
The map $f^1$, defined in the exact sequence (\ref{es1}), is not an isomorphism.
\end{prop}
The proof of such statement will follow from the following:

\begin{cl}
$$H^{1}({\frak gl}_{+}(\infty))^{C}\simeq {\bC}.$$
\end{cl}

\prf 
Let us start to observing that

$$H^{1}({\frak gl}_{+}(\infty))\simeq \big({\frak gl}_{+}(\infty)/[{\frak gl}_{+}(\infty), {\frak gl}_{+}(\infty)]\big)^{'}\simeq {\bC},$$
and identifying $[{\frak gl}_{+}(\infty), {\frak gl}_{+}(\infty)]$ with $\frak sl_{+}(\infty)$,
i.e with the Lie algebra of infinite matrices of finite rank, having trace equal to zero.\\
In particular this implies that the only non trivial class $[\phi]\in H^{1}({\frak gl}_{+}(\infty))$ corresponds to a (closed) cochain $\phi\in C^1({\frak gl}_{+}(\infty))$ whose kernel is $\frak sl_{+}(\infty)$.\\
Let us now define the action of the (abelian) Lie algebra $C$ on $H^{1}({\frak gl}_{+}(\infty))$:
for any $\phi\in C^1({\frak gl}_{+}(\infty))$ and $[Z]\in C\simeq \cL_{L}/{\frak gl}_{+}(\infty)$,
define:
\begin{equation}
([Z].\phi)(\alpha)=\phi([Z+\beta,\alpha]),\label{e20}
\end{equation}
where $Z\in \cL_{L}$ and $\beta\in {\frak gl}_{+}(\infty))$. On the other hand, $\phi$ being a cocycle, we have that:
$$\phi([Z+\beta,\alpha])=\phi([Z,\alpha]).$$
It is a simple calculation to show that $[\cL_{L},{\frak gl}_{+}(\infty)]\subset \frak sl_{+}(\infty)$ so that, from the hypothesis on $\phi$, we conclude that $\phi([Z,\alpha])=0$, i.e:
$$[Z].\phi=0,$$
or that $\phi$ is $C$-invariant.
\endprf

\begin{rk}

We want to remark the following: as in the finite dimensional case, also the Lie algebra $\frak gl_{+}(\infty)$ is not simple. In fact $\frak sl_{+}(\infty)$ is an ideal in $\frak gl_{+}(\infty)$. At the same time we want to underline the difference between the infinite dimensional case and the finite dimensional one. In the finite dimensional case, i.e for a
given $n\in {\bZ}_{>0}$, we have that the quotient ${\frak gl}(n)/{\frak sl}(n)\simeq {\bC}\simeq {\cal Z}({\frak gl}(n))$, where ${\cal Z}({\frak gl}(n))$ is
the center of ${\frak gl}(n)$.

In the infinite dimensional case we still have that ${\frak sl}_{+}(\infty)\subset {\frak gl}_{+}(\infty)$ is an ideal and that
the quotient algebra is one dimensional, but in this case such a quotient does not correspond to any ideal in $\frak gl_{+}(\infty)$. In particular ${\cal Z}({\frak
gl}_{+}(\infty))=\{0\}$.
\end{rk}

\section{Conclusion and Outlook.}
In this paper we gave results about the structure of the the Lie algebra $\cL_{L}$. We first discussed its relevance for the structure of quantum field theory.
Having motivated its study,  we showed that $\cL_{L}$ is the (non-abelian) extension via $\frak gl_{+}(\infty)$ of a commutative Lie algebra. We also showed that
this extension does not split. Furthermore, we described the cohomology of $\cL_{L}$ and proved that the second cohomology group of this Lie algebra is infinite
dimensional, allowing for infinitely many non-equivalent central extensions.

It should be very interesting to understand the physical meaning of the central extensions of this Lie algebra in the future, in particular their relations with
those DSEs. In future work we will study more closely the representation theory of this Lie algebra with the hope to shed some light on these problems.

\vskip20pt
\paragraph{\bf Acknowledgements.}
I.M.~thanks the IHES for hospitality during a stay in May of 2004.  We thank Takashi Kimura for several useful conversations. I.M wants to thank Pavel Etingof
and Victor Kac and Zoran Skoda for stimulating discussions and valuable advices. We thank Claudio Bartocci and Ivan Todorov for carefully reading a preliminary version of the
manuscript and for suggesting several improvements.

\end{document}